\newcommand{\la}{\langle}
\newcommand{\ra}{\rangle}

\newcommand{\beq}{\begin{eqnarray}}
\newcommand{\eeq}{\end{eqnarray}}

\newcommand{\cl}{\centerline}

\newcommand{\btem}{\bibitem}

\documentstyle[12pt]{article}

\begin{document}

\hfill{UTHEP-276}

\hfill{April, 1994}

\vspace{1.5cm}

\vspace{5pt}
 \cl{\Large{\bf Vector  Mesons in Nuclear Medium}}
 \cl{\Large{\bf -- an Effective Lagrangian Approach -- }}

\vspace{1.0cm}
\cl{H. Shiomi\footnote{e-mail address: siomi@nucl.ph.tsukuba.ac.jp}
 and T. Hatsuda\footnote{e-mail address: hatsuda@nucl.ph.tsukuba.ac.jp}}

\vspace{0.5cm}

\noindent
\cl{\em Institute of Physics, University
 of Tsukuba, Tsukuba, Ibaraki 305, Japan}

\vspace{1cm}

\cl{\today}

\vspace{2cm}

\cl{\bf Abstract}
  Effective masses of $\rho$ and $\omega$ mesons
 in nuclear medium are studied in a hadronic effective theory.
  Both  the pole position and the screening mass
 decrease in nuclear matter due to the polarization of the nucleon Dirac sea.
 The physical origin of the decrease is a  reduction of the wave function
 renormalization constant induced by
 the tensor (vector) interaction of the $\rho$ ($\omega$) with the nucleon.
  Relation to the results of the QCD sum rules is also discussed.

\newpage

\setcounter{equation}{0}
\renewcommand{\theequation}{\arabic{equation}}

\noindent
 1. INTRODUCTION.\ \ \
 Effective masses of the vector mesons
 ($\rho$, $\omega$, $\phi$) in nuclear medium \cite{HL,BR}
 have recently attracted wide  interests.
 The decrease of the masses in nuclei is interpreted as
 an evidence of the  partial restoration of chiral symmetry \cite{BDSW}.
 Direct experiments using the  $e^+e^-$ production
 in $\gamma - A$
 and $p-A$ processes are also proposed
  to check the mass shift \cite{shimizu}.\footnote{See also \cite{HKL}
 for the vector mesons in hot matter as a signal of the formation of the quark
 gluon plasma in the relativistic heavy ion collisions.}

As has been  shown by Lee and one of the
 present authors using the QCD sum rules \cite{HL},
 the $\rho$ and $\omega$ masses around the nuclear matter
  density $\rho_0$ can be parametrized as
 \beq
\label{scaling}
m_{\rho,\omega}^*/m_{\rho,\omega} \simeq 1 - (0.18 \pm 0.05) (\rho/\rho_0),
\eeq
 with $m_{\rho,\omega}^*$ ($m_{\rho,\omega}$)  being the pole
 position of the $\rho$ and $\omega$  propagators in the medium
 (vacuum).
 Brown and Rho also predicted similar decrease  on the basis of the
  dilatational symmetry in the chiral lagrangian \cite{BR}.
 In these approaches, the decrease is intimately related to the
 chiral structure of the QCD vacuum in the presence of matter.
 On the other hand,  in the conventional hadronic approaches where
  only the polarization of the nucleon
 Fermi sea is taken into account, the masses of the vector mesons
 stay constant or increase only slightly  \cite{chin,asakawa}.

The purpose of this article is twofold:
 First one is to show explicitly that
 $m_{\rho,\omega}^*$  do decrease
 even in the level of the hadronic effective theory if one properly
 takes into account the fluctuation of the nucleon Dirac sea.
 We will also analyze the physical origin of the decrease.
 Another purpose is to clarify the difference between
  the ``real mass'' defined
 by the pole position of the propagator and the
 ``screening mass'' defined by the damping
 in the space like region of the propagator.
  They are equal in the vacuum but different in the medium, and
  the distinction between the two should be made carefully when
 one analyses experiments with different kinematics.
 So far, the effect of the Dirac sea has been studied only for
 the real mass of the $\omega$-meson \cite{KS,TBA,CJ,PW}.
   As for $\rho$, it is not obvious whether the real and screening masses
  decrease
 with the same manner as $m^*_{\omega}$ in hadronic models:
 In fact, the $\omega N N$ interaction is dominated
 by the vector coupling,  while
   the $\rho N N$ interaction is dominated
 by the tensor coupling as is known from
  nucleon form factors and the nuclear forces \cite{mac}.
 We will show that the $\rho NN$ tensor coupling
 actually plays a crucial role for  $m^*_{\rho}$.

\vspace{0.5cm}

\noindent
2. LAGRANGIAN. \ \ \
Let's start with an  interaction lagrangian of $\rho,\omega$ with
 the nucleon:
\beq
\label{interaction}
L_{int} =g_\alpha \left [ \overline{{\psi}} \gamma_\mu \tau^a {\psi}
-\frac{\kappa_\alpha}{2M} \overline{{\psi}} \sigma_{\mu \nu} \tau^a {\psi}
\partial^\nu \right ] V^\mu_a , \hspace{0.5cm} \alpha=\{\rho, \omega\}\ \ ,
\eeq
where $a$ runs from 0 through 3, $V_0$ ($V_{1-3}$) corresponds to
 the $\omega$ ($\rho$) field, $\tau^a$ is the isospin matrix
 with $\tau^0$=1, and $M$ is the nucleon mass.
 The numerical values of the coupling constants
  ($g_{\alpha}$, $\kappa_{\alpha}$)  will be given in section 4.

In the one-loop level, the density dependent part of the
 self-energy  comes only from  the nucleon-loop\footnote{The self interaction
 of the $\rho$ meson gives density dependence only from two or higher
  loops.
  The coupling of $\rho$
 with in-medium pions analyzed in \cite{asakawa}
 is also  the higher loop effect and will not be  considered in this paper.}:
\beq
\label{polarization}
\Pi_{\mu \nu}^{ab}(q)=-\frac{i}{(2\pi)^4}\int d^4k
{\rm Tr}[\Gamma_\mu^a G(k+q) \Gamma_\nu^b G(k)]  \ \ ,
\eeq
where $(a,b)$ are the isospin indices and
\beq
\label{vertex}
\Gamma_\mu^a=g_\alpha [\gamma_\mu \tau^a -
\frac{\kappa_\alpha}{2M}\sigma_{\mu \lambda} iq^\lambda \tau^a], \hspace{1cm}
\Gamma_\nu^b=g_\alpha [\gamma_\nu \tau^b +
\frac{\kappa_\alpha}{2M}\sigma_{\nu \lambda} iq^\lambda \tau^b]\ \ .
\eeq
The nucleon propagator in the medium
 $G(k)=G_F + G_D$ reads
\beq
\label{propagatorf}
        G_F &=&\frac{\gamma \cdot k^* +M^*}{{k^*}^2-{M^*}^2+i\epsilon},\\
\label{propagatord}
        G_D &=&i\frac{\pi}{E(k)}(\gamma \cdot k^*+M^*)
          \delta(k^*_0-E(k)) \theta(k_F-|{\bf k}|)\ \ .
\eeq
 Here $k^{*\mu}\equiv (k^0-g_\omega \la V^0 \ra , {\bf k})$
 with $\la \cdot \ra $ being the ground state expectation value,
 $ E(k)= \sqrt{{\bf k}^2+ {M^*}^2}$ and  $k_F$
 is  the fermi momentum.
 As for the effective nucleon mass $M^*$,
  we adopt
the result of the relativistic Hartree approximation
 with vacuum fluctuation \cite{chin}
 which is shown in Fig. 1.
 $\Pi_{\mu \nu} $  in (\ref{polarization}) is composed
 of two parts $\Pi_{\mu \nu} = \Pi^{0F}_{\mu \nu} + \Pi^D_{\mu \nu}$:
 the first term corresponds to the
 fluctuation of the Dirac sea of the nucleons with mass $M^*$,
  while the second term
  corresponds to the fluctuation of the
  Fermi sea and the Pauli blocking.  $\Pi^{0F}_{\mu \nu}$
  generally  has divergences to be subtracted.  We
 will show our subtraction procedure in section 3
  and define $\Pi^{F}_{\mu \nu}$
 as the subtracted polarization.

The vector meson propagator in the medium has a general form
\beq
\label{mesonpropagator}
D_{\mu \nu} = {-P_L \over q^2-m^2 + \Pi_L}
              + {-P_T \over q^2-m^2 + \Pi_T},
\eeq
where we have suppressed isospin indices $(a,b)$.
 $m$ denotes the $\rho$ or $\omega$ mass in the vacuum
 and
 $P_{T}$ ($P_{L}$) is the projection
operator \footnote{$P_T^{\mu \nu} =
  g^{\mu i} (g_{ij} + k_i k_j/{\bf k}^2)g^{\nu j}$ and
 $P_L^{\mu \nu} = e^{\mu}e^{\nu}$ with
 $e^{\mu}= {i \over \sqrt{k^2}}(|{\bf k}|,{k_0 \over |{\bf k}|}{\bf k})$.}
 to the
 transverse (longitudinal) direction to
  ${\bf k}$.
  $\Pi_{T,L}$ is related to $\Pi_{\mu \nu}$ as
 $\Pi_L = - (q^2/{\bf q}^2) \Pi_{00},
\Pi_T = (\Pi_l^l + (q_0^2/{\bf q}^2) \Pi_{00})/2$.
To  obtain (\ref{mesonpropagator}), we have used the Steukelberg propagator
 with $\lambda \rightarrow \infty$ \cite{IZ} as a free propagator of the
 massive vector mesons.

\vspace{0.5cm}

\noindent
3. SUBTRACTION PROCEDURE.\ \ \
The interaction (\ref{interaction}) is not  renormalizable
  in the conventional sense. This does not, however, cause essential
 difficulties,
  since the requirement of the strict renormalizability
 is not necessary in effective theories.  A typical example is
  the non-linear $\sigma$ model
 as a low energy effective theory of QCD. The model contains
 infinite series of the higher dimensional operators
 which play a role to cancel the divergences
 emerging from the loops of the lower dimensional operators \cite{WEIN}.
  In this letter, instead of developing a systematic subtraction
 procedure, we
 will take a phenomenological way to extract $\Pi_{\mu \nu}^F$
 from $\Pi^{0F}_{\mu \nu}$.  First of all, we are interested only in the
 density dependence of $m_{\rho,\omega}^*$, thus we will subtract away
  both divergent and finite parts coming from the
  nucleon-loop at zero density.
  This corresponds to a set of the renormalization conditions
 $\partial^n \Pi^F(q^2)/\partial (q^2)^n \mid_{q^2=m^2} = 0$
 ($n=0,1,2, \cdot \cdot \cdot \infty $)
 or equivalently
   the  infinite series of counter terms which normalize
  the propagator in the vacuum to  $1/(q^2-m^2)$.
 At finite density, we will adopt similar conditions
 $\partial^n \Pi^F(q^2)/\partial
 (q^2)^n \mid_{M^* \rightarrow M, q^2=m^2} = 0$
  ($n=0,1,2, \cdot \cdot \cdot \infty $).
 This together with a condition that
 the higher dimensional counter terms contain only polynomials
 with respect to the hadron fields, one can
 uniquely single out the density dependent part from $\Pi_{\mu \nu}^{0F}$.
 Although our  procedure is physically plausible
 and does not suffer from the Landau ghost problem \cite{TBA},
 it is still ``a'' way to subtract the divergences among many other
 possibilities.
 For the $\omega$ meson,  our procedure
  is equivalent to that adopted in \cite{CJ}.
  We have also checked that the qualitative results presented in this paper
 are not affected even when we take other subtraction schemes
  given in \cite{KS,TBA,PW}.

 Using the dimensional reguralization and the above subtraction procedure,
  one obtains the following $\Pi_{\mu \nu}$  for the $\rho$-meson.
 ($Q_{\mu \nu} \equiv q_{\mu}q_{\nu}/q^2 - g_{\mu \nu})$.
\beq
\label{result}
\Pi^{ab}_{\mu \nu} & =& \delta^{ab} \left ( Q_{\mu \nu}
   \Pi^F +\Pi^D_{\mu \nu} \right ), \\
\Pi^F & = & \Pi_v^F+\Pi_{v,t}^F+\Pi_t^F, \ \ \
 \Pi^D_{\mu\nu} = \left(\Pi^D_v+\Pi^D_{v,t}+\Pi_t^D \right )_{\mu\nu}, \\
\Pi_v^F&=& \frac{g_\rho^2}{\pi^2} q^2
\int^1_0 dx \ x(1-x)\log \left \{ \frac{{M^*}^2-q^2x(1-x)}{M^2-q^2x(1-x)}
 \right \},\\
\Pi_{v,t}^F &=& (\frac{g_{\rho}^2\kappa_\rho }{2M})\frac{M^* q^2}{\pi^2}
\int^1_0 dx \log \left \{\frac{{M^*}^2-q^2x(1-x)}{M^2-q^2x(1-x)}\right \},\\
\Pi_t^F &=& (\frac{g_{\rho} \kappa_\rho }{2M})^2 \frac{q^2}{2\pi^2}
\int^1_0 dx \{{M^*}^2 +q^2 x(1-x) \}
\log \left \{ \frac{{M^*}^2-q^2x(1-x)}{M^2-q^2x(1-x)}\right \},\\
(\Pi^D_v)_{\mu \nu} &=& g_\rho^2\frac{\Pi^\omega_{\mu \nu}(q)}{g_\omega^2}, \\
(\Pi^D_{v,t})_{\mu \nu}
 &=& Q_{\mu \nu}
 \left(\frac{g_{\rho}^2 \kappa_\rho}{2M}\right)4M^* q^2 I_0(q),\\
(\Pi^D_t)_{\mu \nu} &=& \left (\frac{g_{\rho} \kappa_\rho }{2M}\right)^2 q^2
 \left [ \frac{- \Pi^\omega_{\mu \nu}(q)}{g_\omega^2}
  + Q_{\mu \nu} \left \{(4{M^*}^2+q^2)I_0(q)+ {\rho_\sigma \over {M^*}}
  \right \} \right ] .
\eeq
Since  $ \Pi^\omega_{\mu \nu}(q)$, $I_0(q)$ and
 $\rho_\sigma $ are defined in ref.\cite{KS},  we will not
 recapitulate them here.
 $\Pi_{\mu \nu}$ for the $\omega$ meson is simply
 obtained by the replacement
($g_\rho, \kappa_\rho) \rightarrow (g_\omega, \kappa_\omega)$
 in the above formulas.  We have neglected the mixing
 of $\omega$ with the scalar meson $\sigma$ which does
 not modify our results qualitatively.\footnote{Our $\Pi^{F,D}$ for
  the $\omega$ meson agree with the
 previous calculations \cite{KS,TBA,CJ,PW}
 except for the different subtraction
 procedure.  $\Pi^D$ for the $\rho$ meson agrees with ref.\cite{shakin}
 except that $M^*=M$ is taken and the sign of $\Pi^D_{v,t}$ is
 opposite to ours in \cite{shakin}.
 Our formulas of $\Pi^F$ for the $\rho$ meson are new.}

\vspace{0.5cm}

\noindent
4. COUPLING CONSTANTS. \ \ \
 We will take the two sets of the coupling constants
  given in Table 1.
   $\kappa_{\omega}=0$ is taken in both sets,
 since the $\omega NN$ tensor coupling is generally
 small (e.g. $\kappa_{\omega}=
 0.12$ in the vector dominance model).

\vspace{0.7cm}

\cl{
\begin{tabular}{|c|rr|} \hline
 & set I & set II \\ \hline \hline
$g_{\rho}$      & 2.63 &2.72  \\ \hline
$\kappa_{\rho}$ &  6.0 & 3.7  \\ \hline
$g_{\omega}$    & 10.1 &  10.1   \\ \hline
\end{tabular}
}

\vspace{0.5cm}

\noindent
Table 1: Two different sets of the coupling constants adopted in Fig. 2 and
 Fig. 3.

\vspace{0.5cm}

Set I is obtained from the $N-N$ forward dispersion relation
 \cite{GK}.  The Bonn potential of the $N-N$ force gives
 similar values with this set.
 Set II is obtained by the vector-meson dominance together  with the
 $\rho$ universality \cite{sakurai}. In the latter case, one cannot
 determine the $\omega NN$ coupling, therefore we adopted the same coupling
 with the one in the first set.
 A major difference between the two sets is the strength of the
 $\rho NN$ tensor coupling. (See ref.\cite{mac} for the
 detailed discussion on the vector-meson coupling constants.)
 In our calculations,  vertex form factors
 are not taken into account for simplicity.

\vspace{0.5cm}

\noindent
5. REAL, SCREENING and INVARIANT MASSES. \ \ \
In the following, we will focus on the transverse polarization
 $\Pi_T$ defined in eq. (\ref{mesonpropagator})
 and consider the inverse propagator
\beq
\label{inverse}
D_T^{-1} (q_0,|{\bf q}|) = q^2 - m^2 +
  \Pi_T^D(q_0,|{\bf q}|) + \Pi_T^F(q^2) \ \ .
\eeq
 Detailed account including the discussion on $\Pi_L$
  will be given elsewhere \cite{SH}.

 Let us define three kinds of masses $m_{re}^*$ (real mass),
 $m_{inv}^*$ (invariant mass) and $m_{sc}^*$ (screening mass).
 $m_{re}^*$ is defined as a lowest zero of $D_T^{-1}(q_0, 0)$.
 It is the quantity to be compared with that in the QCD sum rules (eq.(1))
 and is related to the  peak position of the $e^+e^-$ spectrum obtained
 from the decays of vector mesons in nuclei \cite{shimizu}.
 The invariant mass $m_{inv}^*$ is defined as a
  lowest zero of $D_T^{-1}$ with $\Pi_T^D$ neglected, in which case
  $D_T^{-1}$ is a function of $q^2$ only.
 $m_{inv}^*$ here contains only the fluctuation of the Dirac sea
 by definition.
 Since the rotational invariance leads to the equality
 $\Pi_T^D (q_0,0) = \Pi_L^D(q_0,0)$, $m_{re}^*$ and $m_{inv}^*$
 turn out to be the common poles in both longitudinal and transverse
 part of the vector meson propagator.
 Finally, we define the screening mass $m_{sc}^*$
 as a pure imaginary zero  of $D_T^{-1}(0,|{\bf q}|)$.
   If there is such a pole at $|{\bf q}|=im_{sc}^*$,
 it  contributes to the meson propagator in the coordinate space
 as  $D_T (t=0, {\bf x} \rightarrow \infty)
 \sim \exp (-m_{sc}^*|{\bf x}|)$.\footnote{Strictly speaking,
 $D_T(0,|{\bf q}|)$ has also cuts in the complex
 $|{\bf q}|$ plane which give
 rise to the Friedel oscillations in $D_T(t=0,{\bf x} \rightarrow \infty)$.
  We will not consider such contribution in this paper.}
   In general,  $m_{sc}^*$  for the transverse propagator
 takes different value
 from that in the longitudinal one.

If we neglect the Fermi sea polarization $\Pi^D$,
 one has a simple relation between the invariant mass  $m_{inv}^*$
 and the  finite  wave-function renormalization $Z$ in medium.
 Since $\Pi_T^F(=\Pi_L^F)$ is proportional to $q^2$ as can be seen
 from eq.(9)-(12),
 the meson propagator near the mass shell is written as
\beq
\label{Zfactor}
D_{\mu \nu} \sim {1 \over {Z^{-1}q^2 - m^2}} = {Z \over {q^2 - Zm^2}} \ \ ,
\eeq
which leads to  $m_{inv}^* = \sqrt{Z} m$.   $Z$ is nothing but
 the probability to find  physical $\rho (\omega)$ in the vacuum
 inside the physical  $\rho (\omega)$ in the medium.
 Note that $Z$ can be larger or smaller than unity depending on the
 sign of $M^*-M$.

\vspace{0.5cm}

\noindent
6. NUMERICAL RESULTS AND DISCUSSIONS. \ \ \
 In Fig. 1, the effective masses of $\omega$ are shown together with
 $M^*/M$. The dashed line  denotes $m_{inv}^*/m$.
 One sees that
  $m_{re}^*$, $m_{inv}^*$ and  $m_{sc}^*$ all decrease at finite density,
 e.g. $m_{re}^*/m \simeq 0.8$ at $\rho=\rho_0$.
 By comparing $m_{re}^*$ with $m_{inv}^*$,
 one also observes that there are two competing effects:
 (a)  polarization of the  Dirac sea of the nucleons with $M^*$
  which tends to decrease $m_{inv}^*$, and (b)
 polarization of the Fermi sea and the  Pauli blocking which
 contributes positively to
  $m_{re}^*$.
  We found that (a) dominates over (b).
 Our   result for $m_{re}^*$
 is also  consistent with that in the previous analyses \cite{KS,TBA,CJ,PW}.

 There is a physical reason why $m_{inv}^*$ decreases in the medium:
 Since $M^* < M$ in nuclear matter, it is easier for vector mesons to
 dissociate into the $N\bar{N}$ pair in the medium than in the vacuum.
 In other words, physical $\omega$ is more dressed by the $N \bar{N}$
 pairs in the medium, which leads to
 $Z < 1$ and $m_{inv}^*/m=\sqrt{Z} < 1$.

In Fig. 2 (Fig. 3), we have shown the effective masses
 of the $\rho$ meson with the parameter set I (II).
 The strong $\rho NN$ tensor coupling plays a dominant role
  and gives
 $m_{re}^*/m \simeq 0.6 - 0.7$ at $\rho =\rho_0$.  The
 polarization of the Dirac sea is again the most important ingredient
 and the suppression of $Z$
 is the main reason for the mass reduction.
 Note also that $m_{sc}^* < m_{inv}^*$ for the $\rho$-meson,
  which is opposite to
 the $\omega$-meson case.

 It is in order here to make final  remarks:

\noindent
(i)  The  reduction of $m_{re}^*/m$
 is consistent with that in the QCD sum rules (eq.(1))
 for the $\omega$ meson, and even larger reduction is observed
 for the $\rho$-meson in this paper.
  One should, however, remember that we did not consider any
 vertex form factors for the $\rho N N$ and $\omega N N$ couplings.
 Such form factors will attenuate the magnitude of the
 mass shift of the $\rho$-meson and the $\omega$-meson in a different way.
 From Fig. 1-3,  one
  also observes considerable non-linearity of $m_{re}^*$
 as a function of density.  This is contrast to the linear
 dependence in eq.(1). Further study is necessary to
 clarify the origin of this difference.

\noindent
 (ii) $m_{re}^*$ has direct relevance to the production of
 lepton pairs as we have mentioned \cite{shimizu,HKL}.
 On the other hand, $m_{sc}^*$ is related to the $t$-channel
 exchange of the vector mesons in nuclear processes such as
 $ K^{+}-{}^{12}{\rm C} $ scattering \cite{BDSW}.
 In this case, however, one should also take into account
 the reduction of $Z$ in (\ref{Zfactor}) which partly
 cancels the effect of the mass reduction as shown in \cite{CJ}.

\noindent
 (iii)  Despite some quantitative differences between the
 result of the effective theory here
 and that in the QCD sum rules,
  the physical origin of the decreasing
 $m_{re}^*$ is quite similar in two
 approaches.  The driving forces of the mass reduction are the
   fluctuation of the Dirac sea in the effective theory
 and the change of the chiral condensate $\la (\bar{q}q)^2 \ra$
 in the QCD sum rules.  Physically they are both
 related to the structure of the QCD vacuum
 in nuclear matter. On the other hand,
  the fluctuation of the Fermi sea (particle-hole excitations)
  in the effective theory and the twist 2 condensate
  $\la \bar{q} \gamma_{\mu}D_{\nu}q \ra$ in the QCD sum rules
  contribute positively to  $m_{re}^*$. They can be interpreted as
  the scattering of the vector mesons by the
 valence nucleons in the nuclear matter.
  One also finds that Dirac beats Fermi in both approaches.

\vspace{1.0cm}

\noindent
{\bf Acknowledgements}:
We would like to thank
Akihiko Kato and  Toshio Suzuki for useful discussions and
 suggestions.

\newpage

\cl{\bf Figure Captions}

\vspace{1cm}

\noindent
Fig. 1: Effective masses of the nucleon  and the
 $\omega$ meson as a function of the baryon density.
 $m_{re}^*$, $m_{sc}^*$ and $m$ denote
 the real mass, screening mass and the mass in the vacuum, respectively.
 The dashed line corresponds to  the invariant mass in medium
 $m_{inv}^*/m$.

\vspace{0.7cm}

\noindent
Fig. 2: Real, screening and invariant  masses of the $\rho$-meson
 in the  parameter set I. The dashed line corresponds to
 $m_{inv}^*/m$.

\vspace{0.7cm}

\noindent
Fig. 3: Same quantities with Fig. 2 in  the parameter set II.

\end{document}